\documentclass[sigconf,screen]{acmart}

\setcopyright{cc}
\setcctype{by}
\acmYear{2026}
\copyrightyear{2026}
\acmConference[MODELS 2026]
{ACM/IEEE 29th International Conference on Model Driven Engineering Languages and Systems}
{October 4--9, 2026}
{Malaga, Spain}
\acmBooktitle{ACM/IEEE 29th International Conference on Model Driven Engineering Languages and Systems (MODELS '26), October 4--9, 2026, Malaga, Spain}
\received{2026-07-15}
\received[accepted]{2026-07-31}

\usepackage[T1]{fontenc}
\usepackage{listings}
\usepackage{enumitem}
\usepackage{siunitx}
\usepackage{makecell}

\DeclareUrlCommand{\code}{\urlstyle{tt}}

\definecolor{CodeFrame}{RGB}{180,180,180}
\definecolor{CodeBg}{RGB}{248,248,248}
\definecolor{CodeLineNo}{RGB}{120,120,120}
\definecolor{CodeEmph}{RGB}{120,40,160}
\definecolor{CodeKeyword}{RGB}{0,70,140}
\definecolor{CodeString}{RGB}{120,60,0}
\definecolor{CodeComment}{RGB}{100,100,100}

\lstdefinelanguage{schedulemodel}{
  morekeywords={
    generation,
    manipulation,
    dbcFile,
    traceFile,
    every,
    on,
    when,
    do,
    delay,
    drop,
    insert, 
    changeRate,
    set,
    perturb,
    bitflip,
    previous,
    not,
    and,
    or
  },
  sensitive=true,
  morecomment=[l]{\#},
  morestring=[b]",
}

\lstdefinestyle{code-base}{
  basicstyle=\ttfamily\scriptsize,
  columns=fullflexible,
  keepspaces=true,
  showstringspaces=false,
  upquote=true,
  tabsize=2,
  breaklines=true,
  breakatwhitespace=true,
  frame=single,
  framerule=0.6pt,
  rulecolor=\color{CodeFrame},
  backgroundcolor=\color{CodeBg},
  framesep=2pt,
  xleftmargin=0pt,
  xrightmargin=0pt,
  aboveskip=0.5\baselineskip,
  belowskip=0.5\baselineskip,
  numbers=none,
  numberstyle=\scriptsize\color{CodeLineNo},
  numbersep=0.7em,
  stepnumber=1,
  mathescape=false,
  breakindent=0pt,
  keywordstyle=\bfseries\color{CodeKeyword},
  stringstyle=\color{CodeString},
  commentstyle=\itshape\color{CodeComment},
}

\lstdefinestyle{code-color}{
  style=code-base
}

\lstdefinestyle{code-print}{
  style=code-base,
  keywordstyle=\bfseries,
  commentstyle=\itshape,
  stringstyle=,
}

\DeclareSIUnit{\fps}{FPS}

\begin{document}

\title{CANcept: Model-based CAN Traffic Generation and Manipulation}

\author{Lino Wertz}
\orcid{0009-0006-6446-8026}
\affiliation{%
 \institution{Karlsruhe Institute of Technology}
 \country{Germany}
}

\author{Junes Sheikhi}
\orcid{0009-0002-1563-7816}
\affiliation{%
 \institution{Karlsruhe Institute of Technology}
 \country{Germany}
}

\author{Florian Fehrle}
\orcid{0009-0003-7367-580X}
\affiliation{%
 \institution{Karlsruhe Institute of Technology}
 \country{Germany}
}

\author{Adrian Rupp}
\orcid{0009-0007-9724-312X}
\affiliation{%
 \institution{Karlsruhe Institute of Technology}
 \country{Germany}
}

\author{Tianhai Liu}
\authornote{Corresponding author.}
\orcid{0000-0001-5881-1920}
\affiliation{%
 \institution{Karlsruhe Institute of Technology}
 \country{Germany}
}

\author{Philipp Kern}
\orcid{0000-0002-7618-7401}
\affiliation{%
 \institution{Karlsruhe Institute of Technology}
 \country{Germany}
}

\author{Carsten Sinz}
\orcid{0000-0001-9718-1802}
\affiliation{%
 \institution{\mbox{Karlsruhe University of Applied Sciences}}
 \country{Germany}
}
\author{Bernhard Beckert}
\orcid{0000-0002-9672-3291}
\affiliation{%
 \institution{Karlsruhe Institute of Technology}
 \country{Germany}
}

\renewcommand{\shortauthors}{Wertz et al.}
\newcommand{\toolName}{CANcept}

\begin{abstract}
Testing timing-related safety properties of Controller Area Network (CAN)-based software requires precise control over transmitted data and communication timing.
Existing open-source tools typically encode such scenarios in low-level scripts, making them difficult to maintain and evolve.
We present \toolName{}, an open-source model-based tool for specifying, generating, replaying, and manipulating CAN traffic.
A traffic schedule model (TSM) defines message timing and content transformations, and a DBC-based communication model (DCM) defines encoding and decoding between message-level traffic and CAN frames.
\toolName{} combines both models in one execution mechanism for generated and manipulated traffic, including replay of transformed traces.
A preliminary evaluation indicates that \toolName{} realizes the specified scenarios, with low timing deviation and stable execution under elevated traffic rates.
\end{abstract}



\maketitle

\section{Introduction}
\label{sec:introduction}

Controller Area Network (CAN)~\cite{CAN20} buses remain widely used in automotive systems to exchange information among Electronic Control Units (ECUs) due to their efficient broadcast mechanism for real-time in-vehicle communication~\cite{ISO11898-1}.
CAN traffic is transmitted on the bus as raw data link layer frames, which engineers commonly interpret as logical messages containing signals such as engine speed or temperature.\nobreak\footnote{In CAN terminology, a \textit{message} denotes the logical, decoded representation with signal values, whereas a \textit{frame} denotes its encoded binary representation on the bus.}
These messages are typically defined in Database CAN (DBC) files, which specify how message-level signal values are encoded into and decoded from frames~\cite{VectorCANdb}.
DBC files, thus, provide a structural basis for interpreting CAN messages and mapping application-level signals to raw frame bytes.

Regression and robustness testing of CAN-based software requires test scenarios controlling both the content and temporal behavior of CAN traffic.
Besides defining transmitted signal values, such scenarios must specify characteristics such as message periods, ordering, repetition, or absence.
These characteristics are necessary to reproduce expected operating conditions and to exercise timing-dependent faults in interactions between ECUs~\cite{Rimen1999CANFI,Olufowobi2020SAIDuCANT,Song2016TimingIDS}.
For example, an NHTSA recall report~\cite{NHTSA26V104} describes a software defect where a race condition during initial power-up between an integrated trailer module and a CAN standby-control signal could leave the module powered but unable to communicate with the vehicle.
The defect could cause loss of trailer stop lamps, turn signals, and braking functionality, thereby increasing the risk of a crash.

\looseness=-1
Existing open-source CAN tools, such as can-utils~\cite{can-utils} and SavvyCAN~\cite{savvycan}, support functions such as monitoring and interpreting CAN traffic.
However, test scenarios are commonly implemented using low-level scripts in, e.g., Python or Bash.
Maintaining and evolving such tests is challenging as the intended traffic schedule is often implicit and interwoven with low-level implementation details.
Changes may inadvertently preserve obsolete transmission frequencies, modify unrelated timing behavior, or fail to exercise schedule-dependent faults.

We present \toolName{}, an open-source GUI tool for specifying CAN test scenarios and generating the corresponding traffic for timing-aware regression and robustness testing.

Its central abstraction is a \textit{Traffic Schedule Model (TSM)}, 
which explicitly defines message timing, signal values, and traffic manipulations.
\toolName{} combines this model with a \textit{DBC-based Communication Model (DCM)} and SocketCAN-compatible interfaces~\cite{SocketCAN}.
The DCM defines how logical messages and signals are encoded into CAN frames, while the TSM defines when messages occur and how their timing and content evolve.
Together, these models separate test-scenario specifications from low-level encoding and transmission details.
In a typical workflow, the user loads a DCM file, configures a TSM through the graphical editor,
concurrently monitors, records, generates, or manipulates CAN traffic on virtual or physical CAN buses to exercise the software under test.

\toolName{} is model-based because (i) the DCM is the single authoritative representation shared across every stage of the testing workflow, rather than merely being imported for decoding, and (ii) the TSM captures the execution semantics of CAN traffic, rather than only being translated into an opaque sequence of transmission commands. Current implementation constructs the DCM from DBC files, but the model itself is independent of the DBC syntax.

This paper makes the following \textbf{contributions}:
\begin{itemize}[leftmargin=*]
\item We present \toolName{}, an open-source graphical tool for model-based CAN testing.

\item We introduce the TSM to specify CAN traffic timing, content, and manipulation using a concrete graphical syntax.

\item We provide an execution mechanism that combines TSM-defined traffic behavior with DCM-based frame encoding and decoding.

\item We compare \toolName{} with open-source CAN tools and evaluate its execution and scenario-realization accuracy.
\end{itemize}

\section{Foundations and Models}
\label{sec:models}

\toolName{} operates on
a DCM that defines the structure and interpretation of CAN messages, and a TSM that specifies test scenarios with conditional traffic and content transformations.

\subsection{Communication Foundations and Model}
\label{sec:communication-model}

We first formalize the DCM used by \toolName{}, based on a subset of DBC-style CAN message definitions~\cite{VectorCANdb}.
The same model could also be derived from other CAN description formats, such as ARXML~\cite{AUTOSARARXML} and KCD~\cite{KCDFormat}.
This DCM provides the structural foundation for the TSM introduced in Section~\ref{sec:schedule-model}.
Throughout this paper, \textit{message-level} refers to decoded logical messages and signal values, and \textit{frame-level} to their encoded CAN frames and payload bytes.

A DCM consists of a set of message definitions $C=\{m_1,\ldots,m_n\}$, where each message $m_i \in C$ contains a set of signal definitions.
Each \textit{signal definition} is represented as the tuple $s = (\mathrm{name}_s,\mathrm{pos}_s,\mathrm{len}_s,\allowbreak \mathrm{ord}_s,\mathrm{sign}_s,\mathrm{factor}_s,\mathrm{offset}_s,\mathrm{min}_s,\mathrm{max}_s,\mathrm{unit}_s, \mathrm{src}_s,\mathrm{dst}_s)$.
The components $\mathrm{factor}_s$ and $\mathrm{offset}_s$ define the mapping between the message-level value and the frame-level value, namely
\[
    v^{\mathrm{msg}}_{s} = v^{\mathrm{frm}}_{s}\cdot \mathrm{factor}_s + \mathrm{offset}_s \quad \text{and} \quad
    v^{\mathrm{frm}}_{s} = (v^{\mathrm{msg}}_{s}-\mathrm{offset}_s)/\mathrm{factor}_s.
 \]
During signal encoding, $v^{\mathrm{frm}}_{s}$ is represented as a signed or unsigned integer according to $\mathrm{sign}_s$, ordered according to $\mathrm{ord}_s$ (little-endian or big-endian), and written into the payload byte array starting at bit position $\mathrm{pos}_s$.
The field occupies exactly $\mathrm{len}_s$ bits.
Signal decoding reverses this process with the same interpretation.
The components $\mathrm{min}_s$ and $\mathrm{max}_s$ define the admissible range of values.
The optional element $\mathrm{unit}_s$ specifies the physical unit in which the value is reported.
The optional elements $\mathrm{src}_s$ and $\mathrm{dst}_s$ identify the ECUs that provide and consume the signal, respectively.

A \textit{message definition} is represented as
\(
m=(\mathrm{id}_m,\mathrm{name}_m,\mathrm{len}_m,S_m),
\)
where the elements determine how a logical message is serialized into a CAN frame.
CAN operates at the data-link layer: during message encoding, the message-level values of the signals in $S_m$ are converted and packed into the payload byte array according to their signal definitions.
The parameter $\mathrm{len}_m$ fixes the payload length in bytes, and payload bits not assigned to any signal remain unused or are initialized according to the encoding policy.
The identifier $\mathrm{id}_m$ is placed in the CAN frame header and is used for bus arbitration.
The communication endpoints for each signal are given by the contained signal definitions through $\mathrm{src}_s$ and $\mathrm{dst}_s$.

CAN traffic denotes the ordered occurrence of frames on a CAN interface.
Each occurrence consists of a CAN frame and its transmission time.
CAN traffic can be observed on a physical CAN bus or a virtual CAN interface, e.g., using SocketCAN~\cite{SocketCAN}, and recorded as a CAN trace:
$T^{x} = [(o^{x}_1,t_1),\ldots,(o^{x}_n,t_n)]$, $t_i \in \mathbb{R}_{\geq 0}$,
where $o^{x}_i$ denotes the $i$-th occurrence and $t_i$ its timestamp.
The superscript $x\in\{\mathrm{msg},\mathrm{frm}\}$ distinguishes message-level and frame-level representations.
A message-level trace contains message instances with decoded signal values.
A frame-level trace contains frames with raw payload bytes.
Given a DCM, both representations encode the same signal values and can be converted into one another.

\noindent
\textbf{Running Example}
\label{par:example}
We use an engine-status message as a running example throughout the paper.
The following DBC fragment defines the \texttt{EngineStatus} message and its \texttt{rpm} and \texttt{temp} signals:

\begin{lstlisting}[style=code-base]
BO_ 256 EngineStatus: 8 EngineECU
 SG_ rpm  : 0|16@1+ (1,0) [0|65535] "rpm" DashboardECU
 SG_ temp : 16|8@1+ (1,0) [0|255] "degC" DashboardECU
\end{lstlisting}

The message has CAN identifier $256=\texttt{0x100}$, an 8-byte payload, and $\mathit{EngineECU}$ as sender.
It contains an unsigned 16-bit little-endian engine-speed signal starting at bit 0 and an unsigned 8-bit temperature signal starting at bit $16$.
Both signals use factor 1 and offset 0, have ranges $[0,65535]$ and $[0,255]$, respectively, and $\mathit{DashboardECU}$ as receiver.
In our DCM, this yields
$m_{\mathrm{eng}} = (\texttt{0x100}, 8, \{s_{\mathrm{rpm}},s_{\mathrm{temp}}\})$,
with signals $s_{\mathrm{rpm}}$ and $s_{\mathrm{temp}}$ represented directly.
For the running example below, we assume initial values \texttt{rpm}$=3000$ and \texttt{temp}$=95$.
A simple message-level trace in which \texttt{rpm} increases by \SI{500}{RPM} and \texttt{temp} by \SI{2}{\degreeCelsius} every \SI{100}{\milli\second} is $T^{\mathrm{msg}} = $ 
\([(\texttt{ES}(3500,97),\SI{0}{\milli\second}),
 (\texttt{ES}(4000,99),\SI{100}{\milli\second}),
 (\texttt{ES}(4500,101),\SI{200}{\milli\second})]\),
where $\texttt{ES}(r,t)$ abbreviates a message instance with $\texttt{rpm}=r$ and $\texttt{temp}=t$.
The first generated message-level values are encoded to the frame-level payload $\texttt{[AC 0D 61 00 00 00 00 00]}$.

\subsection{Traffic Schedule Model (TSM)}
\label{sec:schedule-model}

\begin{figure}[t]
    \centering
    \scriptsize
    \begingroup

    \newcommand{\nt}[1]{\ensuremath{\langle}\textit{#1}\ensuremath{\rangle}}
    \newcommand{\kw}[1]{\texttt{#1}}
    \newcommand{\rep}[1]{#1\ensuremath{^{*}}}
    \newcommand{\opt}[1]{#1\ensuremath{^{?}}}
    \newcommand{\bnf}{\ensuremath{::=}}
    \newcommand{\alt}{\ensuremath{\mid}}
    \newcommand{\assign}{\ensuremath{\leftarrow}}

    \setlength{\tabcolsep}{1pt}
    \renewcommand{\arraystretch}{1.08}

    \begin{tabular}{@{}l@{\quad}c@{\quad}l@{}}
        \toprule

        \nt{schedule}
        & \bnf
        & \nt{generation} \alt{} \nt{manipulation} \\

        \nt{generation}
        & \bnf
        & \kw{generation} \{ \nt{dbcRef}; \nt{frequency}; \rep{\nt{assign}}; \} \\

        \nt{manipulation}
        & \bnf
        & \kw{manipulation} \{ \opt{\nt{dbcRef}}; \nt{traceRef}; \rep{\nt{rule}}; \} \\

        \nt{dbcRef}
        & \bnf
        & \kw{dbcFile} \nt{file} \\

        \nt{traceRef}
        & \bnf
        & \kw{traceFile} \nt{file} \\

        \nt{frequency}
        & \bnf
        & \kw{every} \nt{duration} \\

        \nt{rule}
        & \bnf
        & \kw{on} \nt{msgRef} \kw{when} \nt{predicate} \kw{do} \nt{action} \\

        \nt{action}
        & \bnf
        & \nt{trafficAction} \alt{} \nt{contentAction} \\

        \nt{trafficAction}
        & \bnf
        & \kw{delay} \nt{duration} \alt{} \kw{drop} \alt{} \kw{insert} \nt{valueExpr}\\

        \nt{contentAction}
        & \bnf
        & \kw{set} \nt{signalRef} \assign{} \nt{valueExpr} \alt{} \kw{bitflip} \nt{valueExpr} \\
        &
        &
          \alt{} \kw{perturb} \nt{signalRef} \nt{valueExpr} \nt{valueExpr} \\

        \nt{assign}
        & \bnf
        & \nt{signalRef} \assign{} \nt{valueExpr} \\

        \nt{valueExpr}
        & \bnf
        & \nt{number} \alt{} \nt{signalRef} \alt{} \nt{func} (\nt{valueExpr}) \\
        &
        &  \alt{} \nt{valueExpr} \nt{arithOp} \nt{valueExpr}\\

        \nt{duration}
        & \bnf
        & \nt{number}\,\nt{timeUnit} \\

        \nt{timeUnit}
        & \bnf
        & \kw{ns} \alt{} \kw{\textmu s} \alt{} \kw{ms} \alt{} \kw{s} \\

        \nt{predicate}
        & \bnf
        & \textit{atomic predicate over signal values or message identifiers} \\

        \nt{func}
        & \bnf
        & \textit{math functions, e.g., sin, cos, log, abs, and square root} \\

        \nt{signalRef}
        & \bnf
        & \textit{message-name}\kw{.}\textit{signal-name} \\

        \nt{msgRef}
        & \bnf
        & \textit{message-name} \\

        \nt{file}
        & \bnf
        & \textit{path to a model file} \\
        
        \nt{number}
        & \bnf
        & \textit{integer} \alt{} \textit{decimal} \\

        \nt{arithOp}
        & \bnf
        & \kw{+} \alt{} \kw{-} \alt{} \kw{*}\\

        \bottomrule
    \end{tabular}
    \endgroup

    \caption{Abstract grammar of the TSM.}
    \Description[Abstract grammar of the TSM.]
    {Abstract grammar of the TSM.}
    \label{fig:traffic-schedule-grammar}
\end{figure}

A TSM is a rule-based model for specifying CAN test scenarios in terms of message occurrences, their timing, ordering, and the signal values they carry.
It defines message transmission patterns and how signal or payload values evolve across occurrences.
These patterns are expressed as conditional rules, each consisting of a condition and an action.
Conditions are predicates over expressions involving constants and current or previous values.
Actions modify either message occurrences or their signal values.
Occurrence actions can delay, drop, reorder, duplicate, or change transmission rates, while content actions can assign, bit-flip, or inject random noise within a specified range to signal and payload values.
The TSM applies to both message-level and frame-level traffic because both share the same temporal structure and can be converted through a DCM.
Figure~\ref{fig:traffic-schedule-grammar} presents the abstract syntax of the TSM.
In the running example in Section~\ref{sec:communication-model}, the trace can be generated from this generation schedule:

\begin{lstlisting}[style=code-base, language=schedulemodel]
generation {
  dbcFile "vehicle.dbc";
  every 100 ms;
  EngineStatus.rpm <- EngineStatus.rpm + 500;
  EngineStatus.temp <- EngineStatus.temp + 2;
}
\end{lstlisting}

The resulting message-level trace can then be manipulated by a separate schedule.
The following manipulation delays \texttt{EngineStatus} occurrences whose engine speed is at least \SI{4500}{RPM} by \SI{20}{\milli\second} and clamps temperature values above \SI{100}{\degreeCelsius} to \SI{100}{\degreeCelsius}:

\begin{lstlisting}[style=code-base, language=schedulemodel]
manipulation {
  dbcFile "vehicle.dbc";
  traceFile "vehicle.trace";
  on EngineStatus when EngineStatus.rpm >= 4500 do delay 20 ms;
  on EngineStatus when EngineStatus.temp > 100  do set EngineStatus.temp <- 100;
}
\end{lstlisting}

Assuming the initial values above, the rules are evaluated independently for each matching occurrence.
The temperature rule changes the third and subsequent occurrences to $100$, and the delay rule shifts them by \SI{20}{\milli\second}.
Consequently, a fragment of the manipulated trace is $T^{\mathrm{msg}}=[(\texttt{ES}(3500,97),\SI{0}{\milli\second}),\allowbreak (\texttt{ES}(4000,99),\SI{100}{\milli\second}),\allowbreak (\texttt{ES}(4500,100),\SI{220}{\milli\second})]$.
The first two occurrences remain unchanged; the third is delayed from \SI{200}{\milli\second} to \SI{220}{\milli\second} and its temperature is clamped from $101$ to $100$.

\looseness=-1
The two schedules together describe a controlled engine scenario in which signal values evolve over time and messages satisfying specified state conditions are deliberately delayed or modified.
Such scenarios can expose faults that arise only from particular combinations of signal values and communication timing.
E.g., this scenario can test whether a receiver correctly handles late messages whose temperature value has been clamped once the engine speed reaches \SI{4500}{RPM}.

\section{Tool Overview}
\label{sec:overview}

\begin{figure}[tb]
\centering
\includegraphics[width=.85\linewidth]{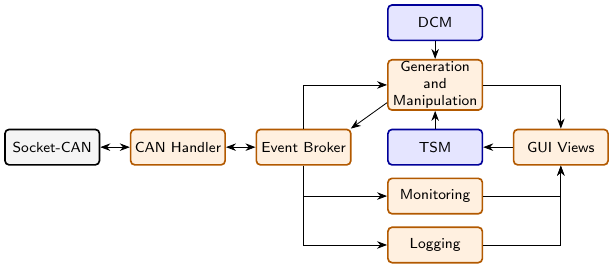}
\caption{Overview of the \toolName{} architecture.}
\Description[Overview of the \toolName{} architecture.]{Overview of the \toolName{} architecture.}
\label{fig:cancept-architecture}
\end{figure}

\toolName{} operationalizes the DCM and TSM as an interactive CAN testing environment.
Figure~\ref{fig:cancept-architecture} shows the tool architecture.
The GUI provides views for DCM inspection, TSM configuration, live DBC traffic monitoring, trace logging, manual transmission, and replay control.
These views exchange commands and updates through an event broker that separates the user interface from the core CAN processing logic.
The core layer combines CAN traffic generation and manipulation based on DCM and TSM, as well as monitoring and logging.
The CAN handler connects the tool to SocketCAN.

\subsection{Traffic Generation and Manipulation}
\toolName{} processes the DCM and TSM through a common model-driven execution mechanism.
Users load both models through the GUI, inspect parsed message, signal, and scheduling information, and configure test scenarios at the model level.
The tool parses each model into an internal representation and binds them by resolving TSM message and signal names against the DCM.
This binding links the two models and could be extended with static analyses that detect inconsistencies between them.

For traffic generation, the execution mechanism evaluates the TSM to create message-level occurrences with concrete timestamps and signal values.
Signal values not assigned by the schedule are initialized using defaults derived from the corresponding DCM definitions.
Each occurrence is then encoded based on the DCM and transmitted or recorded.

For traffic manipulation, recorded frame-level traffic is first decoded through the DCM into message-level occurrences.
The same rule engine then evaluates the TSM over these occurrences.
Traffic actions modify the trace structure, e.g., by delaying, dropping, or reordering messages, and content actions modify signal or payload values.
The resulting occurrences are encoded through the same DCM before transmission, replay, or logging.

The execution mechanism operates on generic representations of messages, signals, expressions, and actions.
Consequently, DCM and TSM instances can be updated by reloading and rebinding without changing the internal model handlers.
Introducing new model constructs or action semantics, however, requires extending the corresponding parser or interpreter.

\subsection{Monitoring and Logging}
\toolName{} also provides monitoring and logging, which are standard capabilities of open-source CAN tools such as python-can~\cite{pythoncan} and SavvyCAN~\cite{savvycan}.
Its GUI displays incoming CAN traffic as time-series views at message and frame levels.
When a matching DCM is loaded, frames are decoded into messages and signal values; otherwise, their identifiers and raw payloads are shown.
The current implementation monitors one CAN interface at a time; concurrent multi-interface support is future work.

Incoming traffic can be recorded as reusable traces in CSV. 
The recorded timestamps, identifiers, payloads, and decoded values can then be inspected offline or loaded back into \toolName{}.
Frame-level traces are decoded using the DCM and can be replayed directly or processed by a TSM.
During unmodified replay, \toolName{} preserves temporal offsets between recorded occurrences, i.e., no schedules are applied.
For manipulation, the TSM modifies the decoded occurrences before they are re-encoded through the DCM.
Thus, recorded traffic can be monitored, logged, manipulated, and replayed within the same execution workflow.

\subsection{Implementation}
\toolName{} is implemented in C++ using Qt~6~\cite{Qt6} and SocketCAN~\cite{SocketCAN}.
SocketCAN allows the tool to operate on both physical and virtual CAN interfaces.
Virtual interfaces support reproducible demonstrations and benchmarks without requiring vehicle hardware.
The current implementation targets Linux-based environments.
Core functional modules communicate through the event broker rather than through direct dependencies.
This design keeps the GUI responsive, isolates time-critical transmission logic, and allows views, analyses, and schedule actions to be added in a modular way.
The abstract grammar in Figure~\ref{fig:traffic-schedule-grammar} defines the principal model elements independently of a concrete serialization format. TSM instances are created in the graphical editor and imported from or exported to JSON, which also stores workflow metadata. The current implementation supports only atomic predicates for signal equality, signal bounds, and message-identifier equality. Textual syntax, a parser, and richer predicates are future work.

\section{Evaluation}
\label{subsec:evaluation}

We evaluate two aspects of \toolName{}: execution-level behavior for standard CAN tasks and realization of declarative TSM-based scenarios.
As open-source baselines, we use python-can~\cite{pythoncan} and SavvyCAN~\cite{savvycan} where comparable functionality is available.
The evaluation comprises two benchmark groups.

The execution-level benchmarks cover four standard CAN tasks: (i) transmission (periodic sending without injected faults), (ii) bit corruption (deterministic payload corruption during transmission), (iii)  raw replay (replay of recorded frame-level traffic), and (iv) DBC-based replay (replay of message-level traffic decoded and re-encoded through the DCM).

The scenario-realization benchmarks cover four test scenario realization tasks: (i) frame loss (probabilistic dropping), (ii) timing violation (threshold-triggered delay), (iii) frame duplication (one-to-one duplication), and (iv) ordering constraint (enforced relative order between messages).

For both benchmarks, we report: (i) frame loss (percentage of scheduled frames not observed), (ii) average jitter (mean absolute timing deviation), (iii) peak jitter (maximum timing deviation), and (iv) throughput (observed frames per second).

Although the experiments run on Linux with a real-time kernel, \toolName{} does not use a dedicated real-time API.
The reported jitter therefore indicates application-level timing stability rather than a bound on hard real-time precision.
In both evaluation results, `n/a` denotes metrics that are not applicable to the tested tool for the corresponding scenario.

\subsection{Execution-Level Comparison}

The evaluation compares standard CAN execution tasks against available baseline tools.
The transmission and bit-corruption benchmarks transmit 60,000 frames at \qty{1,000}{\fps}, while the raw and DBC-based replay benchmarks process recorded traffic for 60 seconds at \qty{10,000}{\fps}.
We additionally report \textit{fault fidelity}, i.e., the percentage of observed fault-injected frames whose payload matches the configured deterministic bit corruption.

\begin{table}
\centering
\caption{Performance results for standard functions.}
\label{tab:evaluation-summary}
\scriptsize
\setlength{\tabcolsep}{1pt}
\resizebox{\linewidth}{!}{%
\begin{tabular}{llrrrr}
\toprule
Scenario & Tool & Frame loss (\%) & Jitter ($\mu$s) & Peak jitter ($\mu$s) & Fault fidelity (\%) \\
\midrule
Transmission & python-can & 0.01 & 219 & 3018 & 100.00 \\
Transmission & SavvyCAN & 10.59 & 431 & 1665 & n/a \\
Transmission & \toolName{} & 0.00 & 2 & 50 & 100.00 \\
\midrule
Bit corruption & python-can & 0.01 & 424 & 3019 & 100.00 \\
Bit corruption & SavvyCAN & 10.52 & 421 & 1870 & n/a \\
Bit corruption & \toolName{} & 0.00 & 2 & 34 & 100.00 \\
\midrule
Replay raw & python-can & 0.00 & 29 & 3722 & n/a \\
Replay raw & \toolName{} & 0.00 & 5 & 841 & n/a \\
\midrule
Replay DBC & python-can & 0.00 & 15 & 2681 & n/a \\
Replay DBC & \toolName{} & 0.00 & 6 & 1334 & n/a \\
\bottomrule
\end{tabular}%
}
\end{table}

Table~\ref{tab:evaluation-summary} summarizes the results.
Fault fidelty is not applicable to SavvyCAN because it lacks deterministic bit corruption.
The results indicate that \toolName{} exhibits substantially more stable timing behavior than the evaluated open-source baselines.
A possible explanation is that \toolName{} uses a dedicated execution pipeline for both traffic generation and replay, whereas other tools rely on more general-purpose scripting mechanisms.

In the transmission and bit-corruption scenarios, \toolName{} achieves \qty{2}{\mu s} jitter with zero frame loss and 100\% fault fidelity.
For replay, \toolName{} again achieves lower average and peak jitter than python-can in both raw and DBC-based replay while preserving frame contents and zero frame loss.
SavvyCAN was not included in the replay comparison because it became unresponsive on the virtual CAN setup.
This indicates a more stable common execution basis for timing-aware generation, fault injection, and replay.

\subsection{Scenario Realization Evaluation}

This evaluation investigates how accurately \toolName{} realizes test scenarios.
Four benchmarks cover frame loss, timing violation, frame duplication, and ordering constraint.

We additionally report \textit{delay error} (absolute deviation from the configured delay), \textit{frame amplification} (ratio of generated frames to original frames), and \textit{ordering violations} (number of occurrences violating the configured order), which directly characterize the intended manipulations.

\begin{table}
\centering
\scriptsize
\caption{Accuracy of scenario realizations.}
\label{tab:manipulation-accuracy}
\setlength{\tabcolsep}{2.5pt}
\renewcommand{\arraystretch}{1.15}
\resizebox{\linewidth}{!}{%
\begin{tabular}{llrrrrrrr}
    \toprule
Scenario & Tool &
\makecell{Loss\\(\%)} &
\makecell{Avg. jit.\\($\mu$s)} &
\makecell{Peak jit.\\($\mu$s)} &
\makecell{Throughput\\(fps)} &
\makecell{Delay err.\\($\mu$s)} &
\makecell{Amp.\\($\times$)} &
\makecell{Order\\vio.} \\
    \midrule
    Frame Loss & python-can
        & 10.18
        & 14
        & 1020
        & 898.17
        & n/a
        & n/a
        & n/a  \\
    Frame Loss & \toolName{}
        & 9.91
        & 1
        & 39
        & 900.95
        & n/a
        & n/a
        & n/a \\
\midrule
    Timing Violation & python-can
        & 0.01
        & 2465
        & 6559
        & 999.92
        & 19
        & n/a
        & n/a  \\
    Timing Violation & \toolName{}
        & 0.01
        & 2465
        & 5013
        & 999.92
        & 15
        & n/a
        & n/a \\
\midrule
    Frame Duplication & python-can
        & 0.00
        & 5
        & 505
        & 18518.51
        & n/a
        & 2.00
        & n/a  \\
    Frame Duplication & \toolName{}
        & 0.00
        & 1
        & 136
        & 18518.52
        & n/a
        & 2.00
        & n/a \\
\midrule
    Ordering Constraint & python-can
        & 0.00
        & 6
        & 1315
        & 10000.01
        & n/a
        & n/a
        & 0  \\
    Ordering Constraint & \toolName{}
        & 0.00
        & 1
        & 198
        & 9999.84
        & n/a
        & n/a
        & 0 \\
    \bottomrule
\end{tabular}
}
\end{table}

Table~\ref{tab:manipulation-accuracy} shows that neither tool violates the configured order or loses frames in the time-independent scenarios.
In the time-dependent scenarios, \toolName{} achieves consistently lower jitter than python-can while meeting the configured delay with slightly smaller error (\qty{15}{\micro\second} versus \qty{19}{\micro\second}).
We configured the scenarios in \toolName{} via the GUI in about \textit{one minute}; equivalent python-can scripts required about \textit{one engineer-hour} and roughly 150 lines of code.
This suggests that \toolName{} reduces manual effort while maintaining accurate scenario realization.

\section{Related Work}
\label{sec:related-work}

\begin{table}
\centering
\caption{Comparison with existing CAN tools.}
\label{tab}
\resizebox{\linewidth}{!}{%
\begin{tabular}{lcccccc}
\toprule
Tool & GUI & DBC & Send/Replay & Timing support & Schedule model & Open source \\
\midrule
can-utils~\cite{can-utils} & no & no & yes & trace-based & no & yes \\
python-can~\cite{pythoncan}/cantools~\cite{cantools} & no & yes & script & periodic/script & no & yes \\
SavvyCAN~\cite{savvycan} & yes & yes & yes & trace-based & no & yes \\
CANoe~\cite{canoe} & yes & yes & yes & yes & proprietary & no \\
\textbf{\toolName{}} & yes & yes & yes & yes & yes & yes \\
\bottomrule
\end{tabular}%
}
\end{table}

Table~\ref{tab} compares tools for testing CAN-based software systems.
Can-utils~\cite{can-utils} provides lightweight frame-level sending, recording, generation, and replay.
Python-can~\cite{pythoncan} and cantools~\cite{cantools} provide scriptable CAN access and DBC-based encoding and decoding.
SavvyCAN~\cite{savvycan} supports capture, visualization, replay, reverse engineering, and fuzzing.
Kayak~\cite{kayak} offers monitoring with XML-based bus definitions.
BUSMASTER~\cite{busmaster} supports simulation and testing, and Scapy automotive extensions~\cite{scapyAutomotive} enable programmable protocol manipulation.
These tools are useful building blocks, but timing and message mutations must generally be encoded in scripts or traces rather than in a reusable scenario model.

Research on CAN fault injection and timing-based intrusion detection highlights the importance of temporal behavior.
Rim'{e}n and Christmansson~\cite{Rimen1999CANFI} consider omitted, delayed, and modified CAN frames, while timing-based approaches~\cite{Olufowobi2020SAIDuCANT,Song2016TimingIDS}
analyze message timing to detect anomalous traffic.
These works motivate explicit modeling of temporal behavior
However, they target specific fault-injection or anomaly-detection techniques rather than a reusable model for specifying and generating CAN test scenarios.

\looseness=-1
Commercial environments such as CANoe~\cite{canoe} provide comprehensive mechanisms for simulation and testing, including programmable timing behavior through simulation nodes and test modules, but lack a comparable open, declarative specification language for jointly describing traffic generation, signal evolution, temporal dependencies, and manipulations.
\toolName{} provides such a language and its underlying model as part of an open-source toolchain.

Unlike tools that mainly use DBC files for encoding and decoding, \toolName{} derives a DCM from a DBC file and uses it as a shared semantic artifact throughout monitoring, replay, logging, generation, and manipulation. The TSM complements the DCM by capturing temporal and dynamic aspects not represented in structural communication descriptions, such as timing relationships, signal evolution, and dynamic triggers.

\section{Conclusion}
\label{sec:conclusion}

We presented \toolName{}, an open-source tool for model-based CAN traffic generation, replay, monitoring, and manipulation.
Its core idea is to combine a DCM with a TSM in one execution mechanism for generated and manipulated traffic.
This lets users inspect, configure, and execute CAN test scenarios at the model level rather than through ad hoc scripts and raw-frame processing.
Recorded traces can be decoded via the DCM, transformed via the TSM, and re-encoded for replay or export, while generated traffic follows the same path.
Our evaluation indicates that \toolName{} realizes the specified loss, delay, duplication, and ordering behaviors with low timing deviation and stable execution under elevated traffic rates.

\begin{acks}
    This work was partially supported by the German Research Foundation (DFG) – SFB 1608 – 501798263.
\end{acks}

\section*{Data Availability Statement}
All prototype and evaluation artifacts are publicly available at~\cite{cancept2026}, the source code is publicly available on GitHub~\url{https://github.com/CANcept/CANcept}.

\bibliographystyle{ACM-Reference-Format}
\bibliography{references}

\end{document}